\documentclass[
]{ceurart}

\usepackage{listings}
\usepackage{multirow}
\usepackage{cleveref}
\usepackage{amsthm}
\usepackage[utf8]{inputenc}
\usepackage{glossaries}
\usepackage{csquotes}

\usepackage[T1]{fontenc}
\usepackage[english]{babel}
\usepackage{microtype} % optional, for aesthetics
\usepackage{tabularx} % nice to have
\usepackage{graphicx}
\usepackage{amsmath,amssymb,amsfonts}
\usepackage{stmaryrd}
\usepackage{textcomp}
\usepackage{xcolor}
\usepackage{mathtools}
\usepackage{pdfpages}

\usepackage{algorithm}
\usepackage{algpseudocode}

\usepackage{listings}
\newenvironment{ownfigure}[0]%
{\begin{figure}[htb!]}%\stepcounter{table}}%
{\end{figure}}

\usepackage{color}
\definecolor{DarkRed}{rgb}{0.75,0,0}
\definecolor{Lightgreen}{rgb}{0.588,1.0,0.588}
\definecolor{DarkGreen}{rgb}{0,0.5,0}
    
\lstdefinelanguage{MontiArc}[]{Java}{
  morekeywords={component, port, in, out, inv, package, import, connect, autoconnect}
}

\lstdefinelanguage{myJava}[]{Java}{
  commentstyle=\color{DarkGreen}\itshape 
}

\lstdefinelanguage{MontiArcAutomaton}[]{Java}{
  morekeywords={component, port, in, out, inv, package, import, connect,
  autoconnect, automaton, state, ocl, java, initial, final,
  noCompletion, chaosCompletion, var, mode, activate, transitions,
  modetransitions}, commentstyle=\color{DarkGreen}\itshape }

\lstdefinelanguage{MCConfig} { 
    morekeywords={config, Require, Model} 
}

\lstdefinelanguage{Manifest} { 
    morekeywords={Manifest, Bundle, ManifestVersion, Name, SymbolicName,
      Version, Require
    } 
}

\lstdefinelanguage{mcGrammar}[]{}{
  morekeywords={
    grammar, package, path, parser, lexer, nows, noslcomments, nomlcomments, 
    noident, nostring, noanything, nocharvocabulary, dotident, identrule,
    xmlcomments, hashcomments, texcomments, freemarkercomments, concept, 
    globalnaming, define, usage, options, true, false, protected, ident, 
    compilationunit
  }
}

\lstdefinelanguage{mcLng}[]{}{
  morekeywords={
    dsltool, language, package, path, parser, root, parsingworkflow, 
    rootfactory, lexer, nows, noslcomments, nomlcomments, noident, nostring,
    dotident, concept, globalnaming, define, usage, options, true, false, 
    protected, ident
  }
}

\lstdefinelanguage{mcManifest}[]{}{
  morekeywords={
    bundle, Bundle, Name, SymbolicName, true, false, Main, Class, 
    Version, Activator, Localization, Require, 
    Exclude, Eclipse, LazyStart, Vendor, Export, Package, 
    ClassPath
  }
}

\lstdefinelanguage{Alloy}[]{Java}{
commentstyle=\color{DarkGreen}\itshape,
  morekeywords={abstract,sig,->,fact,pred,fun,run,for,iff,
  not,no,one,all,some,lone,\#,set,in,and,or,but,exactly,none,univ,Int,assert,check},
  otherkeywords = {[2]????},
    morekeywords = {[2]????},
    keywordstyle={[2]\color{blue}},
    otherkeywords = {[3]????,<,<->,->, &, |, =, !=, !,<:,~},
    morekeywords = {[3]????,<,<->,->, &, |, =, !=, !,<:,~},
    keywordstyle={[3]\color{blue}}
  }

\lstdefinelanguage{mccd}[]{Java}{
  morekeywords={classdiagram,abstract,<<singleton>>,class,int,String,
  association,composition,extends}
}

\lstdefinelanguage{FreeMarker}[]{}{
  keywordsprefix={\#},
  keywords={in},
  commentstyle=\color{DarkGreen}\itshape }

\lstdefinelanguage{Mona}[]{}{
  morekeywords={ex0,all0,ex1,all1,ex2,all2,var0,var1,var2,pred,in,notin,include,union,inter,empty,assert},
  morecomment=[l]{\#},
  commentstyle=\color{DarkGreen}\itshape,
  otherkeywords = {[2]????,next,boolean,init,case,esac},
  morekeywords = {[2]????,next,boolean,init,case,esac},
  otherkeywords = {[3]????,<,<=>,=>, &, |, =, !=, !},
  morekeywords = {[3]????,<,<=>,=>, &, |, =, !=, !},
}

\lstdefinelanguage{myPython}[]{Python}{
  morekeywords={assert},
  morecomment=[l]{\#},
  commentstyle=\color{DarkGreen}\itshape,
}

\lstdefinelanguage{GeneratorConfiguration}[]{Java} {
  morekeywords={
    template, 
    generator, 
    ast, 
    runtime},
}

\lstdefinelanguage{ApplicationConfiguration}[]{Java} {
  morekeywords={
    application,
    behaviors,
    bindings,
    classdiagrams,
    components,
    factories,
    generators,
    map,
    to},
}

\lstdefinelanguage{Isabelle}[]{} {
    morekeywords={
        datatype,
        typedef},
}

\lstdefinelanguage{bpmn}[]{Java} {
    morekeywords={process,event,task,start,end},
}
\usepackage{xspace}
\usepackage{ifdraft}

\newcommand{\no}{\xmark\xspace}

\newcommand{\iow}{I/O$^\omega$\xspace}

\newcommand*{\ie}{\textit{i.e.,}\@\xspace}
\newcommand*{\eg}{\textit{e.g.,}\@\xspace}

\makeatletter
\newcommand*{\etc}{%
  \@ifnextchar{.}%
  {\textit{etc}}%
  {\textit{etc.}\@\xspace}%
}
\makeatother

\newcommand{\xmark}{\ding{55}}%
\definecolor{se-green}{RGB}{0,128,0}
\definecolor{se-blue} {RGB}{0,0,204}

\newcommand{\code}[1]{\texttt{#1}}

\newcounter{requirement}[section]

\usepackage{amsthm}
\theoremstyle{definition}

\glsdisablehyper % No links!

\newacronym{cc}{C\&C}{Component and Connector}

\newacronym{dsl}{DSL}{Domain Specific Language}

\newacronym{mda}{MDA}{Model-Driven Architecture}

\newacronym{cpu}{CPU}{Central Processing Unit}

\newacronym{FDDT}{FDDT}{formal description and development technique}

\newacronym{AFWME}{AFWME}{Apply Fluid with Mechanical Energy}

\newacronym{SRV}{SRV}{Set Rotational Velocity}

\newacronym{SD}{SD}{Synchronous Drive}

\newacronym{SI}{SI}{International System of Units (Système International d'unités)}

\newacronym{OMG}{OMG}{Object Management Group}

\newacronym{V\string&V}{V\string&V}{Validation and Verification}

\newacronym{TEE2ME}{TEE2ME}{Transform Electrical Energy to Mechanical Energy}

\newacronym{MDE}{MDE}{Model Driven Engineering}

\newacronym{ERS}{ERS}{Entity Relationship Schemata}

\newacronym{OCL}{OCL}{Object Constraint Language}

\newacronym{UML}{UML}{Unified Modeling Language}

\newacronym{UML/P}{UML/P}{\gls{UML}/P}

\newacronym{CD}{CD}{Class Diagram}

\newacronym{FD}{FD}{Feature Diagram}

\newacronym{OD}{OD}{Object Diagram}

\newacronym{OM}{OM}{Object Model}

\newacronym{AST}{AST}{Abstract Syntax Tree}
\def\ast{\gls{AST}\xspace}

\newacronym{SMT}{SMT}{Satisfiability Modulo Theories}

\newacronym{CPS}{CPS}{Cyber-Physical System}

\newacronym{cpf}{CPF}{Cyber-Physical Function}

\newacronym{ME}{ME}{Mechanical Engineering}

\newacronym{SE}{SE}{Software Engineering}

\newacronym{PDP}{PDP}{Product Development Process}

\def\sl{\textit{Simulink}\xspace}

\newacronym{OCL/P}{OCL/P}{OCL/Programmable}

\newacronym{sysml}{SysML}{Systems Modeling Language}

\newacronym{xml}{XML}{Extensible Markup Language}

\newacronym{ad}{AD}{Activity Diagram}

\newacronym{sc}{SC}{Statechart}
\def\sc{\gls{sc}\xspace}

\newacronym{ucd}{UCD}{Use Case Diagram}

\newacronym{uc}{UC}{Use Case}

\newacronym{sd}{SD}{Sequence Diagram}

\newacronym{bdd}{BDD}{Block Definition Diagram}

\newacronym{ibd}{IBD}{Internal Block Diagram}

\newacronym{mde}{MDE}{Model-Driven Engineering}

\newacronym{CAD}{CAD}{Computer-Aided Design}

\newacronym{SysML4FMArch}{SysML4FMArch}{SysML for Functional Mechanical 
Architectures}

\usepackage{ifdraft}
\usepackage{ifthen}

\newboolean{debug}
\setboolean{debug}{false}
\ifthenelse{\boolean{debug}}{%
	\usepackage{todonotes}                 %% activate todos
	\newcommand{\xynote}[2]{\todo[inline]{#1: #2}}
	\newcommand{\note}[1]{{\color{blue}\textit{#1}}}
	
	\usepackage{hyperref}                  %% activate colored links
	\usepackage{setspace}                  %% Increase line spacing

}{%
	\newcommand{\note}[1]{}
	\newcommand{\xynote}[2]{}
	
	\hypersetup{hidelinks}                 %% deactivate colored links
}

\begin{document}

%\author{Johanna Grahl}
%\author{Bernhard Rumpe}
%\author{Max Stachon}
%\author{Sebastian Stüber}

\copyrightyear{2022}
\copyrightclause{Copyright for this paper by its authors.
  Use permitted under Creative Commons License Attribution 4.0
  International (CC BY 4.0).}

\conference{arxiv.com}

\title{Tool-Assisted Conformance Checking to Reference Process Models}

\author[1]{Bernhard Rumpe}[%
orcid=0000-0002-2147-1966,
email=rumpe@se.rwth-aachen.de,
url=https://se-rwth.de,
]
\fnmark[1]
\address[1]{Software Engineering, RWTH Aachen University, Germany}

\author[1]{Max Stachon}[%
orcid=0000-0002-6328-3816,
email=stachon@se-rwth.de,
]
\fnmark[1]

\author[1]{Sebastian Stüber}[%
orcid=0000-0002-6636-9375,
email=stueber@se-rwth.de,
]
\fnmark[1]

\author[1]{Valdes Voufo}[%
email=valdes.voufo@rwth-aachen.de
]
\fnmark[1]

\fntext[1]{These authors contributed equally.}

\begin{abstract}
  % Motivation
  Reference models convey best practices and standards crucial for maintaining quality and consistency in various processes. 
  Conformance checks are necessary to ensure adherence to these models and the established guidelines and principles they encode.
  % Research Question
  This paper explores automated conformance checking for concrete process models against reference models using causal dependency analysis of tasks and events.
  % Why Unanswered
  Existing notions of conformance checking for process models focus on verifying process execution traces and lack the expressiveness and automation needed for semantic model comparison, leaving this question unresolved.
  %
  % Big New Idea
  Our approach is integrated into a broader semantic framework for defining reference model conformance.
  % Contribution
  We outline an algorithm for reference process model conformance checking, evaluate its implementation through a case study, and discuss its strengths and limitations.
  % Key Impact
  Our research provides a tool-assisted solution enhancing accuracy and flexibility in process model conformance verification.
\end{abstract}

\begin{keywords}
  Reference Models \sep 
  Process Models \sep 
  Semantic \sep 
  Conformance \sep 
  BPMN \sep 
  Model-Driven Engineering
\end{keywords}

\maketitle

	% !TeX spellcheck = en_US
\section{Introduction} \label{sec:introduction}

Process models play a pivotal role across various domains by providing structured representations of workflows and operations. 
They are instrumental in standardizing processes~\cite{ungan2006standardization,Allweyer2016BPMN2}, optimizing performance~\cite{kougka2018many}, and ensuring compliance with industry standards~\cite{knuplesch2013ensuring,lim2012workflow}. 
As organizations increasingly rely on these models to guide their operational practices~\cite{fernandes2021role}, the demand for robust mechanisms to verify the adherence of concrete implementations to reference models has become paramount.

Reference models serve as authoritative benchmarks that encapsulate standardized processes and best practices~\cite{ITU-T.X.200,GHJV97}. 
In the realm of process modeling, they provide a critical framework against which concrete process implementations can be rigorously evaluated. 
Despite their significance, there is a notable lack of formal conformance verification methods tailored to facilitate appropriate comparisons between specific process models and more general reference models. 
Current approaches to conformance checking in process modeling often fall short in addressing the nuanced causal dependencies inherent in both reference and concrete models. 
Instead, they predominantly focus on verifying execution logs~\cite{RvdA06,vdA12,BMS16,DSMB19}, thereby neglecting the essential aspect of model-to-model comparison.

Conformance checks~\cite{KRS+24} are essential for verifying that concrete process models align with their corresponding reference models, ensuring adherence to best practices and facilitating the timely identification of deviations. 
This alignment is vital for organizations seeking compliance with regulatory standards, optimizing operational efficiency, and maintaining quality assurance. 
However, conformance checking presents significant challenges due to the inherent complexity and variability of real-world processes. 
For instance, the dynamic nature of business operations can lead to frequent changes in process structures, complicating validation efforts. 
Existing methods of semantic model comparison often struggle with limited expressiveness and scalability, which can significantly hinder effective validation in diverse and complex environments~\cite{MRR11a,MRR11b,KR18,RRS23,RSSV24}.

This paper addresses these limitations by introducing an innovative algorithmic approach for tool-assisted conformance checking that leverages causal relations analysis to enhance the expressiveness and automation of the validation process. 
Our motivation arises from the pressing need for more accurate and reliable conformance checks, particularly in scenarios involving intricate process structures and dependencies.

The objective of this study is to develop a novel method that systematically analyzes causal dependencies within reference process models and compares them against corresponding elements in concrete models. 
By doing so, we aim to provide a more flexible and comprehensive solution for conformance verification, thereby advancing the state of the art in process model validation.

\paragraph{Contribution}
\begin{itemize}
    \item Semantic concept for reference process models and conformance
    \item Abstract description of a conformance checking algorithm
    \item Publicly available Java implementation for conformance checking
    \item Evaluation of tool on multiple examples and discussion of results 
\end{itemize}

\paragraph{Structure} In \Cref{sec:motivating}, we present a motivating example that is used to illustrate key concepts. 
\Cref{sec:relatedWork} provides background information and discusses related work in the field. 
We then introduce an abstract description of our conformance checking algorithm in \Cref{sec:concept} and address its complexity, soundness, and completeness. 
Following this, \Cref{sec:implementation} details the implementation aspects. 
The evaluation of our tool is presented in \Cref{sec:evaluation}. 
In \Cref{sec:discussion}, we discuss precision and limitations of our approach, as well as potential threads to validity. 
Finally, we conclude with a summary of our findings and suggest directions for future work.
    \section{Motivating Example} \label{sec:motivating}

\begin{figure*}[ht]
    \centering
    \includegraphics[width=\textwidth]{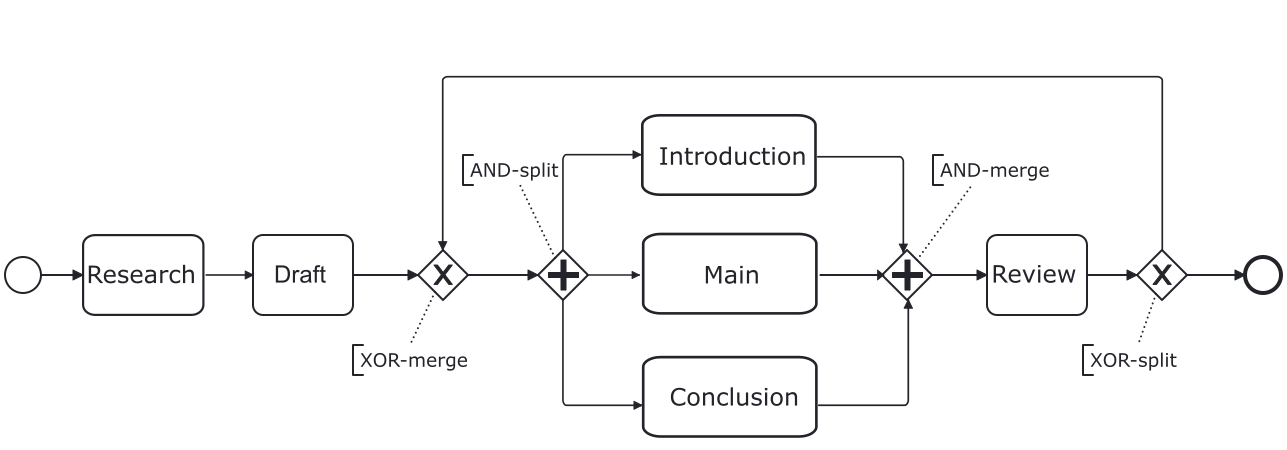}
    \caption{Reference process model for scientific writing.}
    \label{fig:ref}
\end{figure*}

\begin{figure*}[ht]
    \centering
    \includegraphics[width=\textwidth]{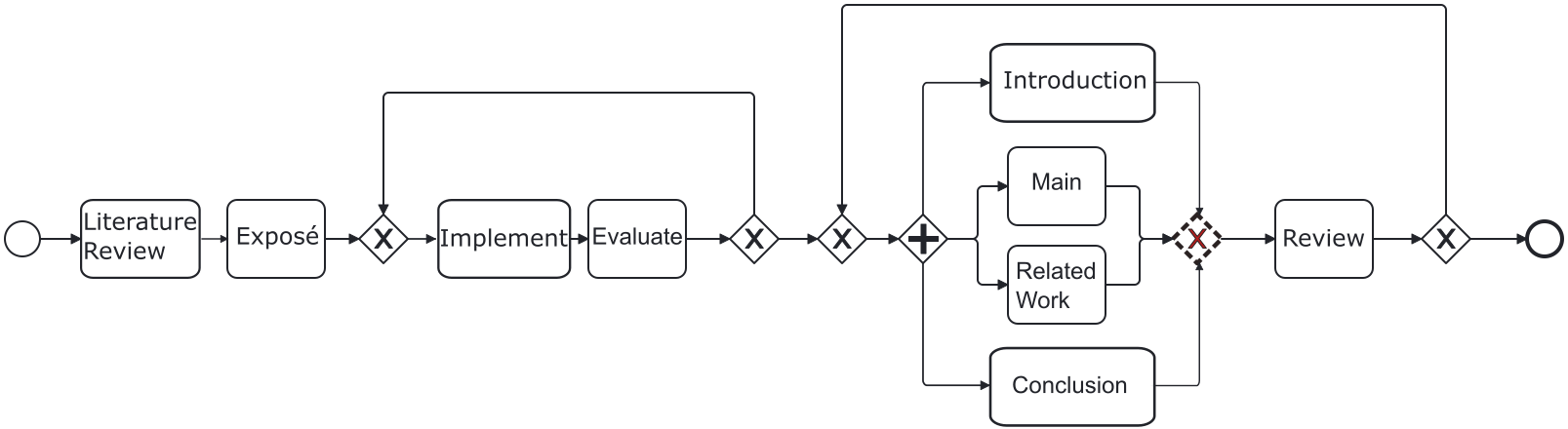}
    \caption{Concrete process model for writing a thesis; the fifth gateway is incorrect and leads to non-conformance.}
    \label{fig:thesis}
\end{figure*}

Consider the process model displayed in \cref{fig:ref}, it specifies a reference process for scientific writing.
After starting the process, the first task is \code{Research}.
After \code{Research} is completed, the next task is to write an initial \code{Draft}.
Following that \code{Introduction}, \code{Main}, and \code{Conclusion} have to be completed.
These three can be worked on in parallel.
Next up is the task \code{Review}, after which the process either ends or the tasks \code{Introduction}, \code{Main}, and \code{Conclusion} are repeated.

Based on this reference model, we want to develop a more refined version of the model for the concrete case of writing a scientific paper for a conference.
This model is displayed in \cref{fig:thesis}:
The thesis starts with a \code{Literature Review} instead of \code{Research}, followed by writing an \code{Exposé} instead of a \code{Draft}.
Afterwards, the tasks \code{Implement} and then \code{Evaluation} have to be completed.
Both may have to be repeated.
Following that \code{Introduction}, \code{Main}, \code{Conclusion}, as well as \code{Related Work} have to be written.
After a \code{Review} of the work these tasks may have to be repeated, as well.
Otherwise, the process ends.

This is the intended specification of the process.
However, on a closer look at the model displayed in \cref{fig:thesis} one might discover a mistake made by the modeler:
The fifth gateway is an \code{XOR-merge} instead of the intended \code{AND-merge}.
As a consequence, the execution behavior of the model does not correspond to the reference model, \ie its semantics was not properly preserved.

A mistake such as this becomes harder to discover the larger the model grows, increasing chances that the model will be implemented and executed when the corresponding software system is deployed.
The severity of the impact such a mistake has can vary, but if compliance to the reference model is legally required, it is best to avoid this situation in the first place.

This mistake in particular does not cause a deadlock, and can therefore not be discovered by a deadlock-analysis of the model.
It might potentially be discovered during testing; after all, merging sequence flows that are executed in parallel might be considered an anti-pattern and corresponding tests might exist.
However, tests would have to be implemented for every well-known anti-pattern and even then there are cases were a mistake does not produce an anti-pattern but still alters the behavior of the concrete model in an unintended way, \eg accidentally placing tasks in a sequence flow in the wrong order.
Instead, to detect mistakes such as this that lead to non-conformance with a reference model, we propose the use of a tool-assisted conformance check.

    \section{Background and Related Work} \label{sec:relatedWork}

In the following we outline necessary background information and discuss related work.

\subsection*{Process Modeling with BPMN}
Business Process Model and Notation (BPMN)~\cite{OMG} is an internationally recognized standard developed by the Object Management Group (OMG). 
Its primary aim is to furnish a notation for business process design that is easily comprehensible to all stakeholders across various levels of the development process, from initial drafts to implementation, and extending to the management and monitoring of these processes. 
BPMN serves as a synthesis of best practices from the business modeling community, delineating the notation and semantics for collaboration, process, and choreography diagrams.

In this paper, we concentrate on the process design aspects of BPMN, specifically addressing the order and interdependencies of task executions within a business process. 
Consequently, we restrict our analysis to a subset of the BPMN syntax and features. Our focus encompasses process definitions that involve \emph{events}, \emph{tasks}, \emph{gateways}, and \emph{sequence flows} that interconnect these elements. 
For the purposes of our study, we categorize gateways into two types: \emph{split} and \emph{merge}. 
Each gateway can further be classified as an \code{AND} gateway, an \code{XOR} gateway, or an \code{OR} gateway.

According to the BPMN standard, the following execution semantics apply to gateways: 
All sequence flows following an \code{AND}-split gateway are executed in parallel. 
In contrast, after an \code{XOR}-split gateway, only one of the subsequent sequence flows is executed. 
\code{OR}-split gateways permit the parallel execution of multiple subsequent sequence flows. 
An \code{AND}-merge waits for the completion of all preceding sequence flows, while an \code{XOR}-merge requires the completion of just one preceding sequence flow. 
Notably, other active flows are not terminated and may continue to pass through the gateway. Lastly, an \code{OR}-merge awaits the completion of all active preceding sequence flows.

\subsection*{Conformance to a Reference Model}

In the context of process mining the terms \emph{conformance checking} or \emph{conformance testing}
usually refer to a comparison of actual process execution recorded in event logs with the behavior specified by a process model, \ie the intended or required procedure.
This is accomplished by aligning traces from the event log with sequences allowed by the process model, and identifying where the real executions deviate from the model specifications.

While traditional conformance checking focuses on the alignment of actual process executions with a predefined model~\cite{RvdA06,vdA12,BMS16,DSMB19,rafiei2019mining,schuster2020scalable,rafiei2025federated}, our investigation shifts towards a theoretical framework that evaluates the structural and semantic alignment between two models: one abstract and one concrete. 
In this paper, we are concerned with a different kind of conformance relation, which pertains to model-to-model conformance involving a reference model and a concrete model. 
Here, the focus is not on whether a process execution trace represents a legal instance of a model but rather on whether all legal executions of the concrete model conform to the reference model.

A reference model is defined as an abstract representation within a given modeling language that captures domain concepts and their domain-specific relations, specifying properties that must hold for any conformant concrete model, \eg which model elements may or must exist and how they interrelate~\cite{KMR24}. 
A robust conformance relation must guarantee that all relevant semantic properties are preserved by conformant concrete models~\citet{KRS+24}. More specifically, a conformant concrete model must contain appropriate incarnations of relevant elements from the reference model. These incarnations are concrete model elements that correspond to reference elements while preserving their interrelations in the concrete model.

To ensure this formally, we require semantic refinement of the reference model by the concrete model in the context of incarnation. This process is crucial to maintain the intended semantics of the reference model and ensure integrity in the relationships among model elements.

Incarnations are specified via an incarnation mapping—a formal specification that defines how elements in the reference model correspond to elements in the concrete model—or automatically derived by a conformance checking algorithm. 
Conformance checking algorithms implement these conformance relations and have been developed for various modeling paradigms, including class diagrams, feature models, and statecharts~\cite{KRS+24}. 
These algorithms facilitate the evaluation of how well a concrete model adheres to its reference counterpart and determine conformance violations occurring in the concrete model.

The incarnation mappings for these algorithms were either specified using stereotypes~\cite{GogollaStereotype,HKR21}, which annotate model elements with additional semantics, or through custom mapping languages that provide a flexible way to define relationships between model elements. 
In cases where explicit mappings are incomplete or ambiguous, name-equality serves as a fall-back option, allowing the algorithm to infer correspondences based on the similarity of element names.

In this paper, we present a novel conformance checking algorithm specifically designed for process models. 
Our approach leverages stereotypes for encoding the incarnation mapping while also employing name-equality as a fall-back option to enhance the robustness of the conformance checking process.

\subsection*{Semantic Differencing}

A denotational semantics definition $sem : M \rightarrow 2^D$ assigns each syntactically correct model of a modeling language $M$ to a set of valid instances within a well-defined and comprehensible semantic domain $D$~\cite{HR04}. 
In this context, refinement is characterized as a subset relation between sets of instances; specifically, a model $A$ is said to refine a model $B$ if and only if $sem(A) \subseteq sem(B)$.
For process models, the valid instances correspond to process execution traces.

To enable concrete models to extend a reference model in a meaningful manner, it is generally more appropriate to adopt an open-world assumption regarding model semantics for conformance relations, rather than a closed-world assumption. 
This perspective allows a concrete model to incorporate additional elements that do not have counterparts in the reference model while still maintaining conformance. 
In terms of process model semantics, this implies that the execution trace of the concrete model may include additional tasks and events, provided that the causal relations among tasks and events in the reference model are preserved by their respective incarnations. 
It is essential to note that we assume tasks and events are uniquely identified by names within a process model, ensuring that each task and event appears only once.

One effective approach to establish a conformance relation is through semantic differencing~\cite{MRR10}, which analyzes the differences between two models based on their legal instances. Various semantic differencing operators have been developed for multiple modeling languages~\cite{MRR11b,MRR11d,LMK14,BKRW17,KR18,BKRW19,DEKR19,DKMR19,Kau21,RRS23,RSSV24}. The conformance checker for feature diagrams~\cite{KRS+24} builds upon a prior semantic differencing approach~\cite{DEKR19} by integrating incarnation mappings.

One approach to formalizing the execution semantics of process models, particularly Business Process Model and Notation (BPMN), is to translate them into Petri nets~\cite{DDO08}. 
However, comparing Petri nets based on their execution traces is generally undecidable~\cite{Hir93}. 
Consequently, existing semantic differencing operators for process models~\cite{MRR11d, KR18} convert activity diagrams into state machines, where the state space is represented by the power set of activities, and the transition function encodes all potential process execution steps, namely, transitions from one set of active tasks to another.
In this context, \cite{MRR11d} employs this translation to perform bisimulation of the models, while \cite{KR18} utilizes language inclusion checking algorithms for finite word automata.

However, a significant drawback of the translation to state machines is the exponential growth of the state space compared to the original activity diagram. 
This power-set automaton construction, which leads to scalability issues, is necessary to capture the semantics of concurrent activities, which must be interleaved in all possible configurations. 
Alternative, more sophisticated semantic models for concurrency, such as partially ordered multisets (pomsets) and Mazurkiewicz traces~\cite{pratt1986modeling, mazurkiewicz1986trace, bloom1991trade}, exist. 
These models mitigate the interleaving explosion problem by focusing on partial orders or equivalence classes of sequences that preserve only causal dependencies and independence. 
Nonetheless, they introduce additional mathematical complexity and necessitate specialized algorithms for determining behavioral equivalences, such as concurrency-aware bisimulation.

Another aspect that complicates the use of bisimulation for conformance checking is the handling of multiple and composite incarnations of reference tasks and events. 
In this context, multiple incarnations occur when a single element in the reference model is incarnated multiple times in the concrete model, while composite incarnations involve a configuration where a single reference element is represented by multiple elements in the concrete model, or vice-versa. 
For instance, in process models, multiple incarnations could manifest as inclusive or exclusive alternatives of tasks or events, whereas composite incarnations would necessitate the parallel or sequential execution of these tasks and/or events.

The challenge lies in the requirement that bisimulation establishes a one-to-one correspondence between states in the reference and concrete models. This requirement becomes problematic when dealing with multiple and composite incarnations, as they introduce a level of complexity that cannot be easily reconciled with the strict nature of bisimulation. 
Additionally, due to our open-world assumption regarding process model semantics, extra tasks and events may be added to the concrete model, which complicates the conformance checking process further. 
Unlike these additional elements, multiple and composite incarnations cannot simply be deleted or ignored in the concrete model, as they are integral to its structure and behavior.

Our approach circumvents the need for full bisimulation by focusing exclusively on local causal dependencies, accepting a trade-off in completeness. 
Initially, we construct two propositional formulas for each task and event in the reference model: one formula captures the causal dependencies to its direct predecessor tasks and events, while the other addresses the dependencies to its direct successors. 
Subsequently, we verify whether these dependencies are maintained for the corresponding incarnations in the concrete model.

    \section{Conformance Checking Approach} \label{sec:concept}

\begin{figure*}[h]
    \centering
    \includegraphics[width=\textwidth]{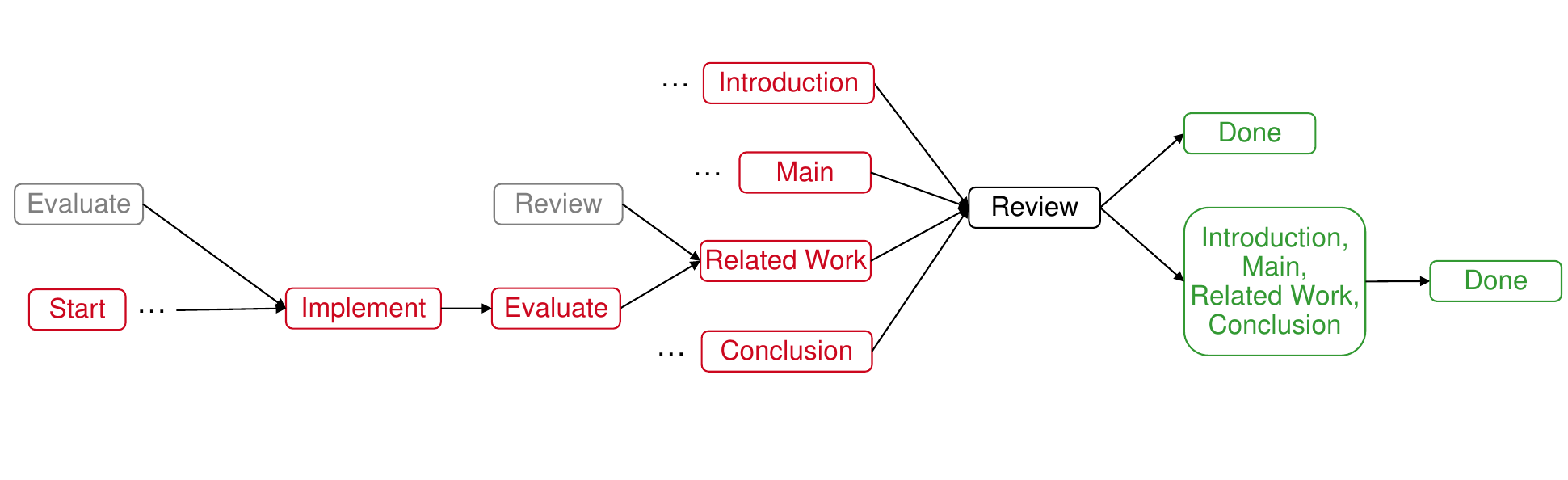}
    \caption{Abbreviated search tree for forward and backward search for the task \code{Review} in the concrete model in \cref{fig:thesis} with satisfying branches in green, non-satisfying in red, and branches deleted due to loops / idleness in grey.}
    \label{fig:search-tree}
\end{figure*}

We define the open-world semantics of a process model as the collection of process execution traces that uphold the local causal dependencies of tasks and events as specified by the process model. In this context, each instance of a task or event is permitted to occur only after a suitable configuration of its predecessor instances has taken place and before an appropriate configuration of its successor instances is realized. Furthermore, these configurations of predecessor and successor instances must also be maintained between instances of the same task or event. In other words, loops that disrupt the local dependency structure of tasks and events are explicitly disallowed.

Our conformance checking algorithm operates under the assumption that an incarnation mapping exists, which correlates each incarnation in the concrete process model to its corresponding task or event in the reference model. Notably, a reference task or event may have multiple incarnations. We analyze each reference task and its incarnations individually, scrutinizing their local causal dependencies with respect to both their predecessor and successor tasks and events as encoded in the graph structure and gateways of the reference process model. This analysis is then juxtaposed with the causal dependencies of the corresponding incarnations in the concrete model.

This approach bears similarity to our conformance check for class diagrams as described in \cite{KRS+24}, where an incarnation is deemed conformant to its corresponding reference element if its properties and relationships with other elements are preserved. However, in the context of concrete process models, it is inadequate to limit our examination to neighboring tasks and events. For instance, consider the following scenarios:

\begin{itemize}
\item \textbf{Insertion of New Tasks or Events:} A new task or event may be introduced between two incarnations of subsequent reference elements. This situation reflects a refinement in accordance with our open-world semantics, as it allows for the evolution of the process model without violating causal dependencies.
\item \textbf{Sequential Execution of Parallel Tasks:} In the concrete model, the incarnations of two parallel tasks might need to be executed in a sequential manner. This represents a refinement since it alters the execution order while still preserving the underlying dependencies defined in the reference model.  
\item \textbf{Violation of Exclusive Alternatives:} The causal relations defined by exclusive alternatives among reference tasks may be contravened in the concrete model if their incarnations are executed sequentially. In such cases, the concrete model fails to be a refinement of the reference model and thus does not conform to it.
\end{itemize}

\subsection*{Algorithm}

The conformance checking algorithm is divided into two distinct phases.
In the first phase, we identify the local causal dependencies of tasks and events within the reference model and encode these dependencies into propositional logic formulas—one for all direct predecessors and another for all direct successors. 
In the second phase, we perform the conformance check on the concrete model by analyzing the causal relation of each incarnation of a task or event with their predecessors and successors in the concrete model, subsequently comparing these relations to the local causal dependencies expressed in the corresponding formulas.

\subsubsection*{Phase 1: Construct the Formula}

In this phase, we compute the direct predecessors and successors of a task or event in the reference model and represent their interrelations as formulas in propositional logic, treating each task and each event as a Boolean variable. 
This process involves executing a depth-first search both forwards and backwards from the reference element.

During the forward search, we branch out at each split gateway, while in the backward search, we branch out at each merge gateway, continuing until we encounter the first task or event on each branch. 
Each gateway is interpreted as a corresponding logical operation applied to the sub-formulas derived from the branches.
The pseudo-code for the forward direction is presented in \cref{alg:sucpred}.

\begin{algorithm}[H]
\caption{A recursive algorithm for computing the successor formula of $n$}\label{alg:sucpred}
\begin{algorithmic}
\Require $x.\text{suc}$ are the predecessor nodes of $x$

\State \Return \Call{SucForm}{$n.\text{suc}$}

\Function{SucForm}{$x$}
  \If{$x$ is an \code{AND}-split gateway}
    \State \Return $AND\bigl(\{$\Call{SucForm}{$s$}$: s \in x.\text{suc}$\}$\bigr)$
  \ElsIf{$x$ is an \code{XOR}-split gateway}
    \State \Return $XOR\bigl(\{$\Call{SucForm}{$s$}$: s \in x.\text{suc}$\}$\bigr)$
  \ElsIf{$x$ is an \code{OR}-split gateway}
    \State \Return $OR\bigl(\{$\Call{SucForm}{$s$}$: s \in x.\text{suc}$\}$\bigr)$
  \ElsIf{$x$ is an event or task of the reference model}
    \State \Return $x$
  \Else
    \State \Return \Call{SucForm}{$x.suc$}
  \EndIf
\EndFunction
\end{algorithmic}
\end{algorithm}

In the backward direction, we treat \code{XOR}-merge gateways identically to \code{OR}-merge gateways, reflecting the execution semantics of BPMN. 
Specifically, multiple preceding sequence flows can be concurrently active and reach the gateway.
The corresponding pseudo-code for this process can be found in \cref{alg:prepred}.

\begin{algorithm}[H]
\caption{A recursive algorithm for computing the predecessor formula of $n$}\label{alg:prepred}
\begin{algorithmic}
\Require $x.\text{pred}$ are the successor nodes of $x$

\State \Return \Call{PreForm}{$n.\text{pred}$}

\Function{PreForm}{$x$}
  \If{$x$ is an \code{AND}-merge gateway}
    \State \Return $AND\bigl(\{$\Call{PreForm}{$p$} $: p \in x.\text{pred}$\}$\bigr)$
  \ElsIf{$x$ is an \code{XOR}- or \code{OR}-merge gateway}
    \State \Return $OR\bigl(\{$\Call{PreForm}{$p$} $: p \in x.\text{pred}$\}$\bigr)$
  \ElsIf{$x$ is an event or task of the reference model}
    \State \Return $x$
\Else
    \State \Return \Call{PreForm}{$x.pred$}
  \EndIf
\EndFunction
\end{algorithmic}
\end{algorithm}

\paragraph{Example: } Consider the task \code{Draft} in the reference process model from \cref{fig:ref}. 
Its only predecessor is \code{Research}, so the formula for the backwards direction only consists of the Boolean variable of the same name.
For successors, we find \code{Introduction}, \code{Main}, and \code{Conclusion} after an \code{AND}-split gateway, meaning that the formula for the forward direction is: 
\begin{center} 
\code{Introduction AND Main AND Conclusion}
\end{center}
If we consider the task \code{Review}, instead, then we get this same formula but for the backwards direction. 
As for the forward direction, \code{Review} is assigned the formula:
\begin{center}
    \code{(Introduction AND Main AND Conclusion) XOR Done}
\end{center}
with \code{Done} being the end event.

\subsubsection*{Phase 2: Check Conformance}
After constructing the two formulas encoding the local causal dependency of the reference task or event to respectively its direct predecessors and its direct successors, we perform a quasi-simulation of the concrete model—both forward and backwards—using breadth-first-search, starting with the incarnation of the reference element. 
The forward search aims to determine whether the incarnations of the successors can be located in a configuration that satisfies the corresponding formula, while the backward search serves a similar purpose for the predecessors.

In the following, we focus solely on the forward direction, as the backward direction follows a largely analogous approach.
A simplified version of our algorithm for the forward direction is presented as pseudo-code in \cref{alg:conformance}.
We initiate the process with the incarnation $n$ and establish the initial branch $b$.
Each branch comprises a set of visited nodes $N$, a set of active nodes $A$, and a result $r$.
Additionally, we maintain the current execution trace, although this detail is omitted from the pseudo-code for simplicity.

The algorithm proceeds iteratively until no branches with active nodes remain. 
In each iteration, for every event, task, \code{AND}-split gateway, and \code{XOR}- or \code{OR}-merge gateway present in $A$, we perform the following steps to progress:
\begin{enumerate}
    \item Remove the current node from $A$.
    \item Add all successor nodes that are not already included in $N$ to $A$.
    \item Update the result of the branch by checking whether the current set of tasks and events in $N$ satisfies the encoded formula.
\end{enumerate}

Due to the presence of exclusive alternatives, we may encounter situations in which a branch that was previously marked as \emph{conform} no longer satisfies the formula. 
As we will elaborate later, this does not necessarily imply that the execution trace represented by this branch is non-conformant; therefore, we designate its status as \emph{unknown}.

When we encounter an \code{XOR}-split gateway in $A$, we create a new branch for each successor node, updating the sets of visited and active nodes, as well as the corresponding result for each new branch. 
Subsequently, we delete the current branch. 
In a similar manner, if an \code{OR}-split gateway is encountered in $A$, we generate a new branch for each subset of successor nodes and then delete the current branch.

Conversely, if the set of active nodes $A$ contains only \code{AND}-merge gateways, we can progress through one of these gateways, provided that all its predecessor nodes are included in $N$. 
If no such gateway exists, we will advance through another available \code{AND}-merge gateway.

Finally, if no active nodes remain but we have not yet reached an end event or returned to $n$, we delete the branch.

Once all branches with active nodes have been exhausted, we return the set of branches that were not deleted and examine their results. 
If any non-conformant branch exists, we conclude that the incarnation is \emph{not conform}, and we return the execution trace as a diff witness. 
Conversely, if a branch with the status \emph{unknown} exists, the conformance status of the incarnation is designated as \emph{unknown}, and we return the trace as a potential diff witness for manual verification. 
Lastly, if all remaining branches are conformant, we classify the incarnation as \emph{conform}.

\paragraph{Example:} Going back to our previous example, we now consider the concrete process model from \cref{fig:thesis}, where we start the forward search for the task \code{Review}.
We branch out in our search because of the subsequent \code{XOR}-split gateway.
One branch terminates in the next step as we reach the end event \code{Done}.
Having visited \code{Done}, the branch satisfies the successor formula:
\begin{center}
    \code{(Introduction AND Main AND Conclusion) XOR Done}
\end{center}
The other branch finds an \code{AND}-split gateway after the loop and adds the tasks \code{Introduction}, \code{Main}, \code{Related Work} and \code{Conclusion} to its list of visited nodes.
In the next step, we reach the starting point \code{Review} and terminate the search in this branch.
This branch also satisfies the successor formula, having visited:
\begin{center}
    \code{[Introduction, Main, Related Work, Conclusion, Review]}
\end{center}
Since all branches satisfy the formula, the task \code{Review} is conform with regards to its successors.
However, this is not the case with regards to its predecessors.
If we backtrack, we immediately encounter an \code{XOR}-split and branch out.
One of the branches will now, before terminating, visit:
\begin{center}
    \code{[Related Work, Evaluate, Implement, Exposé, Literature Review, Start]}
\end{center}
This does not satisfy the predecessor formula:
\begin{center}
    \code{Introduction AND Main AND Conclusion}
\end{center}
As such, the task \code{Review} is not conform.

The abbreviated search tree for both the forward and backward search is displayed in \cref{fig:search-tree}. 
The nodes contain the tasks and events visited in that step and are colored green if the formula is satisfied in this step, red if not, and gray if an already visited element was visited again and the branch will be ignored.

\begin{algorithm}[H]
\caption{Simplified conformance checking algorithm for an incarnation $n$}\label{alg:conformance}
\begin{algorithmic}[1]
\Require $x.\text{suc}$ are the successor nodes of $x$ and $x.\text{pred}$ the predecessors

\State $\mathcal{B} \gets \{(\emptyset, \{n\}, \text{not conform})\}$
\While{$\exists b = (N,A,r) \in \mathcal{B}$ with $A \neq \emptyset$}
  \ForAll{$b = (N,A,r) \in \mathcal{B}$ with $A \neq \emptyset$}

    % go 1 step forward for most nodes
    \ForAll{$x \in A$}
      \If{$x$ is an event, a task, an \code{AND}-split gateway, or an \code{XOR}- or \code{OR}-merge gateway}
        \State $A \gets (A \backslash \{x\}) \cup (x.\text{suc} \backslash N)$
        \State $N \gets N \cup x.\text{suc}$
        \State \Call{UpdateResult}{$b$}
      \EndIf
    \EndFor

    % branching for or- and xor-splits
    \If{$\exists x \in A: x$ is an \code{XOR}- or \code{OR}-split gateway}
      \If{$x$ is an \code{XOR} gateway}
        \ForAll{$s \in x.$suc}
          \State $b_s \gets (N \cup \{s\}, (A \backslash \{x\}) \cup (\{s\}\backslash N), r)$
          \State \Call{UpdateResult}{$b_s$}
          \State $\mathcal{B} \gets \mathcal{B} \cup \{b_s\}$
        \EndFor
      \ElsIf{$x$ is an \code{OR} gateway}
        \ForAll{$S \subseteq x.$suc with $S \neq \emptyset$}
          \State $b_S \gets (N \cup S, (A\backslash \{x\}) \cup (S\backslash N), r)$
          \State \Call{UpdateResult}{$b_s$}
          \State $\mathcal{B} \gets \mathcal{B} \cup \{b_S\}$
        \EndFor
      \EndIf
      \State $\mathcal{B} \gets \mathcal{B} \backslash b$
      
    \ElsIf{$A \neq \emptyset$}
      
      % handle waiting merge nodes
      \While{$A$ contains only \code{AND}-gateways}
        \If{$\exists m \in A: m.\text{pred} \subseteq N$}
          \State $x \gets m$
        \Else
          \State $x \in A$
        \EndIf
        \State $A \gets (A \backslash \{x\}) \cup (x.\text{suc} \backslash N)$
        \State $N \gets N \cup x.\text{suc}$
        \State \Call{UpdateResult}{$b$}
      \EndWhile
      
    \ElsIf{$n \notin N$ and $N$ contains no end event}
      \State $\mathcal{B} \gets \mathcal{B} \backslash b$
      
    \EndIf 
  \EndFor
\EndWhile

\State \Return $\mathcal{B}$

\Function{UpdatResult}{$b = (N,A,r) $}
  \If{$r =\text{not conform}$ and $N$ satisfies the formula}
    \State $r =\text{conform}$
  \ElsIf{$r =\text{conform}$ and $N$ does not satisfies the formula}
    \State $r =\text{unknown}$
  \EndIf
\EndFunction

\end{algorithmic}
\end{algorithm}

\paragraph{Complexity:}\label{subsec:complexity} If no inclusive \code{OR} gateways are utilized, both parts of the algorithm can be executed in polynomial time. 
Specifically, we first employ depth-first search to construct the formula, followed by breadth-first search for the conformance checking. 
However, the complexity increases exponentially with the number of inclusive decision branches. 
Notably, this represents an improvement over previous approaches that relied on power-set automaton construction~\cite{MRR11d, KR18}, as our method does not require interleaving concurrent tasks and events.

\paragraph{Soundness:}
A conformance relation must ensure semantic refinement in the context of incarnations. 
For process models, this entails that every execution trace of the concrete model must be permissible under the reference model. 
Operating under an open-world assumption allows for extensions of the process in the concrete case, which necessitates that the relative order of elements in a trace corresponds to the order of elements in the reference model.

Preserving causal dependencies for incarnations—as implemented by our algorithm—serves as a sufficient condition for semantic refinement under an open-world assumption. At any point during the execution of the concrete process, it is guaranteed that for each active incarnation, all predecessor incarnations can ultimately be identified through backtracking. Furthermore, a suitable configuration of successor incarnations will eventually be established if the execution is continued. Consequently, the resulting execution trace must be permissible according to the reference model.

\begin{figure}[h]
    \centering
    \includegraphics[width=0.6\textwidth]{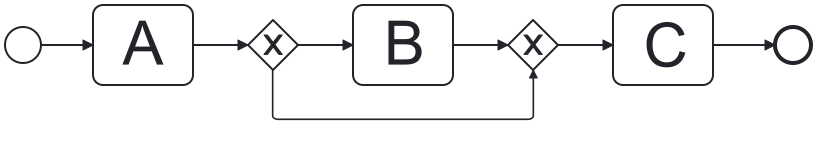}
    \caption{Example of a process model with a skippable task.}
    \label{fig:skip}
\end{figure}

\paragraph{Completeness:} The reduced complexity of our approach, as compared to previous semantic differencing methods for process models~\cite{MRR11a, KR18}, does come at a cost. In certain cases, a branch may reach a configuration that violates the formula concerning an \code{XOR}-constraint after having satisfied the formula in a prior step. This circumstance does not necessarily indicate that the model is non-conformant, as illustrated by the process model depicted in \cref{fig:skip}.

For instance, when checking the conformance of this model against itself, the algorithm first derives the formula \code{B XOR C} for the successors of \code{A}. During the conformance check of \code{A} as an incarnation of itself, the algorithm will branch in the first step due to the \code{XOR}-split and will successfully identify satisfying configurations \code{B} and \code{C}, respectively. However, in the subsequent step, the configuration \code{[B,C]} is encountered, which no longer satisfies the formula. Consequently, the branch is unable to reach a satisfying configuration thereafter.

As a result, the algorithm indicates that it cannot ascertain whether \code{A} is conformant, outputting the configuration \code{[B,C]}. Nevertheless, this task sequence \code{[B,C]} can be extended to form a legal run \code{[start,A,B,C,end]}, as the reference model is identical to the concrete model. This allows for manual verification, confirming that the model is indeed conformant.

    \section{Implementation \& Tool}\label{sec:implementation}

For our implementation, we make use of a textual version of BPMN~\cite{DMM+22} developed with the \code{MontiCore}\footnote{\url{https://monticore.github.io/monticore/}} language workbench~\cite{GKR+06,HKR21}.
To encode the incarnation mapping, we have extended the grammar of the language to allow the annotation of tasks and events with stereotypes~\cite{GogollaStereotype,HKR21}.
Similar to our previous approach for conformance checking of class diagrams~\cite{KRS+24}, incarnations are identified via a stereotype that specifies the name of the incarnation mapping as well as the name of the corresponding reference element.
If for a given reference element no incarnation is specified, a concrete element of the same name is considered as incarnation.

\Cref{lst:wf-example} shows a process model in the textual BPMN syntax that describes a sequential writing process.
The tasks \code{Concept} and \code{Implementation} are each annotated with a stereotype indicating that both tasks are mapped via the incarnation mapping \code{ref} to a task in a reference model with the name \code{Main}.
This process is in fact conform to our reference model for scientific writing displayed in \cref{fig:ref}, since the order of tasks has simply been sequentialized.
\\ \vspace{5mm}

\begin{lstlisting}[language=bpmn,label=lst:wf-example, caption={Example -- BPMN in textual syntax}]
process SequentialWriting {
  event start Start;
  event end Done;

  task Research;
  task Draft;
  task Introduction;
  <<ref="Main">> task Concept;
  <<ref="Main">> task Implementation;
  task Conclusion;
  task Review;

  Start -> Research -> Draft -> Introduction -> Concept -> Implementation -> Conclusion -> Review -> Done;
}
\end{lstlisting}

The conformance-checking tool has been integrated in to the \code{BPMN} language-project of the \code{MontiCore} language-family and is publicly available on GitHub\footnote{\url{https://github.com/MontiCore/bpmn}}.
After building the project the BPMN conformance check can be executed via the \code{BPMN.jar} as demonstrated in \cref{lst:command}.
The tool takes as input a path to the reference model, specified via the option $-r$  and a path to the concrete model, specified via the option $-c$.
Both must be in the form of a \code{.wfm}-file containing the textual specification.
Finally, the name of the incarnation mapping used in the concrete model is specified via the option $-m$.

\begin{center}
\begin{lstlisting}[label=lst:command, caption={Executing the BPMN conformance checker}]
java -jar BPMN.jar                                \
          -i Concrete.wfm -ref Reference.wfm      \
          -m "ref"
\end{lstlisting}
\end{center}

Given the example from \cref{lst:wf-example} as a concrete model and a textual version of our scientific writing process from \cref{fig:ref} as reference model, the conformance check produces the output displayed in \cref{lst:out-conform} on the console.

\begin{center}
\begin{lstlisting}[language=bpmn,label=lst:out-conform, caption={Output in case of conformance}]
Checking Conformance of [Concrete:Sequential] to [Reference:PaperAuthoring]

--- Final Result of Conformance Checking ---
--- All nodes conform to their reference ---
\end{lstlisting}
\end{center}

If instead we consider the thesis-writing process from \cref{fig:thesis} as our concrete model, the tool informs us that the task \code{Review} is not conform with respect to its predecessors. 
It also provides a backtracking sequence from \code{Review} to the start event that demonstrates this, as can be seen in \cref{lst:out-non-conform}. 

\newpage

\begin{center}
\begin{lstlisting}[label=lst:out-non-conform, caption={Output in case of non-conformance}]
Checking Conformance of [Concrete:AntiPattern] to [Reference:PaperAuthoring]

--- Final Result of Conformance Checking ---
The following nodes do not conform: [Review]

-------- Explanations --------: 

Result: Node [AntiPattern:Review] does not conform to Node [PaperAuthoring:Review]
Counter example: The following backtrack [Review, Introduction, Evaluate, Implement, Expose, LiteratureReview, Start] is possible in [AntiPattern] but not in [PaperAuthoring].
\end{lstlisting}
\end{center}

Of course there is also the previously discussed case in which the algorithm is unable to determine conformance or non-conformance.
This happens, \eg, if we check conformance of the process model from \cref{fig:skip} to itself.
In this particular situation, the tool determines that the task \code{A} may not be conform with regards to its successors. 
It then produces a potentially non-conformant sequence from \code{A} to the end event, as can be seen in \cref{lst:out-unknown}.

\begin{center}
\begin{lstlisting}[label=lst:out-unknown, caption={Output in case of potential non-conformance}]
Checking Conformance of [Concrete:Skip] to [Reference:Skip]

--- Final Result of Conformance Checking ---
The status of the following nodes is unknown: [A]

-------- Explanations --------: 

Result: Node [Skip:A] may not conform to Node [Skip:A]
Counter example: The following run [B, C, Done] is possible in [Skip] but may not be possible in [Skip].
\end{lstlisting}
\end{center}

	\section{Evaluation} \label{sec:evaluation}

In order to evaluate the implementation of our algorithm, we constructed a small case study, in which we consider conformant and non-conformant extensions and modifications of reference models using the reference process model for scientific writing displayed in \cref{fig:ref} as our initial model.
Noticeably, we have included ten changes that we consider \emph{conform}, and ten that are non-conform.
We execute the conformance check to compare the modified model to the original and verify the results.
The tests for the case study can be found in the BPMN project at:

\begin{center}
    \code{WorkflowConformance/src/test/java/\\de/monticore/bpmn/conformance/CaseStudyTest.java}
\end{center}
with the corresponding model-files at:

\begin{center}
    \code{WorkflowConformance/src/test/resources/\\de/monticore/bpmn/conformance/caseStudy/}
\end{center}

\subsection{Conformant Modifications}

We consider the following modifications refining, \ie the resulting models should conform to the original model:

\begin{enumerate}
    \item sequentializing parallel tasks
    \item removing a loop
    \item adding new tasks
    \item removing alternatives
    \item parallelizing inclusive alternatives
    \item transforming inclusive into exclusive alternatives
    \item incarnating a task multiple times in parallel
    \item incarnating a task multiple times in sequence
    \item incarnating a task multiple times as inclusive alternatives
    \item incarnating a task multiple times as exclusive alternatives
\end{enumerate}
In all cases the tool informs us that all nodes are conform.

\subsection{Non-Conformant Modifications}

We consider the following modifications non-refining, \ie the resulting models should not conform to the original model:

\begin{enumerate}
    \item switching the order of tasks
    \item removing or not incarnating a task
    \item incarnating a task at a correct and incorrect position
    \item transforming an \code{XOR}-split into an \code{AND}-split
    \item transforming an \code{AND}-split into an \code{XOR}-split
    \item transforming an \code{AND}-merge into an \code{XOR}-merge
    \item transforming an \code{AND}-split into an \code{OR}-split
    \item parallelizing exclusive alternatives
    \item turning exclusive alternatives into inclusive alternatives
    \item sequentializing exclusive alternatives
\end{enumerate}
In all cases the tool identifies the non-conformant nodes and outputs a corresponding diff witness in the form of a run or backtrack sequence.

    \section{Discussion} \label{sec:discussion}

In this study, we explored the concept of conformance checking of concrete models to reference models within process modeling languages such as BPMN. 
We developed an approach that focuses on preserving causal dependencies among tasks and events in the context of their incarnations. 
In this section, we will discuss aspects of uncertainty whithin our approach, its limitations, and potential threats to validity.

\paragraph{Uncertainty:} 
The algorithm we developed ensures that during the execution of a conformant concrete model, incarnations of tasks and events from the reference model can only occur if their requisite predecessor incarnations have been executed beforehand and suitable configurations of their successor incarnations follow.
This design guarantees that the execution trace remains permissible according to the reference model, while operating under the open-world assumption and in the context of incarnation.

In cases where the algorithm determines the non-conformance of an incarnation, it uncovers a path within the model that violates the local dependency of tasks and events as specified by the reference model.
AProvided that the incarnation is reachable and a path to an end event exists, this path can be expanded into a process execution trace that acts as a \emph{diff witness}, \ie a trace that is not permitted by the reference model, while operating under the open-world assumption and in the context of incarnation.

Uncertainty therefore only exists in the form of incarnations whose conformance status cannot be determined and are hence labeled as \emph{unknown}. 
This occurs when a satisfying configuration is followed by a non-satisfying configuration in a subsequent step in the same branch of the forward search, which in turn can only result from visiting incarnations of exclusive alternatives among successors.

Unlike previous conformance checking approaches that exhibit uncertainty regarding semantic refinement~\cite{KRS+24}, our method cannot simply categorize these instances as non-conformant. 
This is because reflexivity is a necessary component of a conformance relation, and as illustrated in \cref{fig:skip}, labeling them as non-conformant would not be justified. 
Consequently, we recognize our approach as somewhat incomplete, necessitating a manual review of these ambiguous cases. 
Fortunately, our algorithm provides both the name of the individual incarnation and the execution trace that is flagged as potentially non-conformant, facilitating this review process.

\paragraph{Limitations:}
A significant limitation of our current approach is that it only considers a subset of BPMN language features~\cite{OMG}. 
Specifically, we focus on basic tasks, events, and the logical gateways \code{XOR}, \code{OR}, and \code{AND}, which together represent a rudimentary foundation for process modeling. 
Furthermore, events are treated similarly to tasks, with our examples restricted to start and end events. 
Future work should explore additional BPMN features and their implications for process semantics.

Certain features, such as lanes, are not integrated into our current trace-based semantics definition and may therefore be overlooked. 
However, elements like expressions—especially executable formal expressions—play a crucial role in defining flow and loop conditions, as well as event triggers. 
Our textual representation of BPMN accommodates such formal expressions, and future enhancements to the algorithm could involve conformance checking of these expressions.

One promising approach would be to utilize SMT-solving to verify the refinement of individual reference expressions against their concrete counterparts. 
Additionally, sub-processes could potentially be managed through a hierarchical conformance check, while call activities may be addressed similarly.

Moreover, data and message flows represent another critical aspect for consideration. 
In our BPMN implementation, usable data types can be defined using class diagrams~\cite{HLN+15,HLN+15a}, allowing us to leverage our existing conformance checks~\cite{KRS+24}. 
If behaviors are specified through operation constraints in OCL~\cite{CKM+02,RSSV24}, operations declared in a class diagram and referenced in a task of the process model could also be subjected to conformance verification.

It is essential to enable a more precise encoding of incarnation mappings, not only to support additional features but also to facilitate more complex incarnations of currently supported features, such as tasks. 
The existing mapping mechanism, which relies on stereotypes, lacks the necessary expressiveness. 
For instance, it does not allow for a task in the reference model to be represented as a sequence of tasks. 
To address these limitations, we propose the development of a custom mapping language that can enhance the expressiveness and versatility of our incarnation mappings.

\paragraph{Threats to Validity:}
Our current interpretation of open-world semantics views any extension of the original model as a refinement that preserves the local causal dependencies between tasks and events, thereby ensuring that our approach is effectively correct by construction. 
However, this definition may not be universally applicable to all scenarios and use cases; alternative interpretations might be more appropriate in certain contexts.

For instance, we explicitly disallow the addition of loops that involve tasks and events present in the model by necessitating suitable configurations of direct successor and predecessor tasks and events between two instances of the same task or event within an execution trace. 
In some scenarios, however, these loops may represent relevant refinement steps. 
It might be sufficient for the aforementioned predecessor and successor configurations to occur just once—respectively, before and after all instances of the task or event within the trace.

To further validate our approach, we plan to enhance our evaluation in the future by identifying relevant example cases from business, industry, and scientific literature to perform a comprehensive analysis.
    \section{Conclusion} \label{sec:conclusion}

This paper presents an innovative approach to conformance checking of reference process models through the introduction of an algorithmic solution that leverages causal relations analysis. 
Our primary motivation was to enhance both the expressiveness and automation of conformance checks, enabling more precise verification of complex process structures. 
We achieved this objective by developing a novel method that systematically examines causal interdependency of elements within reference models and compares it with the interdependency of corresponding elements in concrete models, thereby providing a more adaptable and comprehensive framework for conformance verification.

Our contributions include establishing a semantic concept for reference process models and their conformance, providing an abstract description of a conformance checking algorithm, releasing a publicly accessible Java implementation, and evaluating the tool against multiple examples.
These contributions collectively advance the field by offering a robust methodology and toolset for enhancing conformance checks of reference process models.

Looking ahead, future work will involve conducting industry case studies to validate our approach in real-world scenarios, as well as extending our BPMN feature support to encompass a broader range of elements and language variants ~\cite{CGR09,GR10}. 
These efforts aim to further enhance the practical applicability and robustness of our method, ensuring its relevance across diverse industrial contexts.
    \begin{acknowledgments}
    Funded by the Deutsche Forschungsgemeinschaft (DFG, German Research Foundation) - 250902306
\end{acknowledgments}
	\bibliography{main.bib}

	%\onecolumngrid
		
\end{document}